\documentclass[pra,twocolumn,groupaddress,amsfonts,amssymb,showpacs]{revtex4}

\usepackage{amsmath}
\usepackage{amstext}
\usepackage{latexsym}
\usepackage[dvips]{graphicx}

\newcommand{\beq}{\begin{equation}}
\newcommand{\eeq}{\end{equation}}
\newcommand{\barr}{\begin{eqnarray}}
\newcommand{\earr}{\end{eqnarray}}
\newcommand{\ket}[1]{\left\vert#1\right\rangle}
\newcommand{\Ham}{\mathcal H}

\begin{document}

\title{Information transfer rates in spin quantum channels}

\author{Davide Rossini}
\affiliation{{NEST-CNR-INFM} \& Scuola Normale Superiore,
  	Piazza dei Cavalieri 7, I-56126 Pisa, Italy}

\author{Vittorio Giovannetti}
\affiliation{{NEST-CNR-INFM} \& Scuola Normale Superiore,
  	Piazza dei Cavalieri 7, I-56126 Pisa, Italy}

\author{Rosario Fazio}
\affiliation{International School for Advanced Studies (SISSA)
        via  Beirut 2-4,  I-34014, Trieste,  Italy\\
        NEST-CNR-INFM \& Scuola Normale
        Superiore, Piazza dei Cavalieri 7, I-56126 Pisa, Italy}

\date{\today}

\begin{abstract}

  We analyze the communication efficiency of quantum information transfer 
  along unmodulated spin chains by computing the communication rates of
  various protocols.
  The effects of temporal correlations are discussed,
  showing that they can be exploited to boost the transmission efficiency.

\end{abstract}

\pacs{03.67.Hk, 03.67.-a, 42.50.-p}

\maketitle

\section{Introduction}\label{sec:protocol}
 
Spin networks are an ideal theoretical playground to devise,
test and analyze quantum information processing~\cite{NIELSEN}.
Examples are the implementation of computational models in spin chains
introduced in Refs.~\cite{computation,FITZ} or the possibility to
realize approximate quantum cloners~\cite{DECHIARA}.
In the context of quantum communication, several protocols have been
proposed which would allow distant parties to exchange and/or share
quantum information over chains of permanently coupled
spins~\cite{FITZ,bose03,vittorio05,daniel05,daniel05B,daniel05C,vgdaniel,
OSBORNE,hase05,WOJCIK,KARBACH,PATERNOSTRO,yung,niko,DATTA,DEPASQUALE,Plenio,
SRINIVASA,HE,CAI}.
This research is in part motivated~\cite{bose03} by the need of finding
alternatives to the standard {\em flying qubits} strategies for connecting
regions of high control (i.e. quantum computers).
These latter approaches, in fact, may pose non trivial interfacing problems
with several of the proposed quantum computing architectures
(e.g. ion traps, Josephson junctions, etc.). 
On the other hand, there have been some indications that effective
spin networks could be engineered by using arrays of 
Josephson junctions~\cite{ROMITO}, quantum dots~\cite{IDAMICO},
optical lattices~\cite{OPTICAL} or QED cavities~\cite{CAVITY}.

The ultimate aim of typical communication protocols is to achieve
perfect transfer of any quantum state in a given period of time.
High degrees of efficiency in these protocols also require
the capability of isolating the experimental setup from the external world,
preventing it from decoherence, and to reduce all static internal imperfections.
Two different strategies are usually adopted:
A certain class of schemes achieve efficient communication by tailoring
the system Hamiltonian~\cite{DATTA,KARBACH,PATERNOSTRO,yung,niko,Plenio,SRINIVASA},
with a corresponding minimal (if not null) cost in terms of encoding and
decoding operations.
In the second approach, communication efficiency is obtained through complex
encoding and decoding operations but with minimal requirements on the Hamiltonian of
the chain~\cite{OSBORNE,hase05,daniel05,daniel05B,daniel05C,vgdaniel,HE,WOJCIK}.
Apart from unavoidable experimental noise, which inevitably reduces the fidelity
of the state transfer, even the theoretical analysis of these protocols
is problematic, because of the dispersive nature of the information
propagation~\cite{OSBORNE,bose03}.
Indeed, on one hand, dispersion does not permit a sharp definition of
transmission times, while, on the other hand, it is responsible for the presence
of feedback and memory effects~\cite{MEMORY,MEMORY1} in the quantum communication.

In this paper we analyze the communication performances of some
spin-chain communication protocols.
By studying the asymptotic number of qubits transmitted per second,
we will show that efficient mechanisms of information transfer
can be devised by carefully exploiting the dispersive dynamics of the chain.
The paper is organized as follows. In Sec.~\ref{sec:model}  we introduce
a prototypical spin chain communication model.
In Sec.~\ref{sec:due} we show how one can use the internal dynamics
of the chain to improve the communication efficiency by focusing on the simplest
not-trivial solvable case (i.e. a chain with only two intermediate spins). 
In Sec.~\ref{sec:dualrail} instead we analyze the case of 
arbitrarily long chains: here a lower bound on the 
attainable communication rates is provided
by exploiting the dual-rail protocol of Ref.~\cite{daniel05}.
Finally, in Sec.~\ref{sec:conclusions} we draw our conclusions.

\begin{figure}[t]
  \includegraphics[scale=0.63]{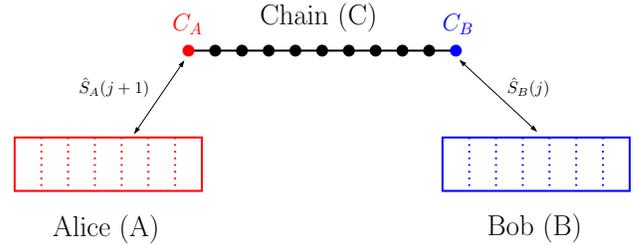}
  \caption{Communication scenario. Alice (A) and
    Bob (B) control their memory, and have access respectively to the
    first and the last qubit of the chain (C). The intermediate spins
    of the chain act as a quantum channel for quantum communication
    between Alice and Bob.
    At regular time intervals both Alice and Bob perform a SWAP gate
    between one of her/his memory qubit and the spin of the
    chain to whom each of them has access.
    During the time between two consecutive swaps the chain is let
    freely evolve, according to the Hamiltonian $\hat{\Ham}$.}
  \label{fig:channel}
\end{figure}

\section{The model}\label{sec:model}

As in Refs.~\cite{bose03,vittorio05,daniel05,daniel05B,daniel05C,vgdaniel,
OSBORNE,hase05,WOJCIK}, the spin communication channel 
we consider here is an array $C$ of $N$ permanently coupled spin $1/2$
that interact through an  Hamiltonian $\hat{\Ham}$. 
The total spin component $\hat{S}_z$ of the system
(e.g. ferromagnetic Heisenberg or $XY$ interaction) is a constant of motion.
The chain $C$ constitutes the physical channel along which
quantum information is transmitted.
As shown in Fig.~\ref{fig:channel}, we assume that the sender (Alice)
and the receiver (Bob) of the messages
have access to two distinct subsets $C_A$ and $C_B$ of the spins
of the chain (typically the first and the last spin) and to two
distinct sets of ancillary qubits (i.e. Alice's memories 
$A\equiv \cdots A_3 \;  A_2 \;  A_1 $ and 
Bob's memories $B\equiv B_1 \; B_2 \; B_3 \;\cdots $, respectively).
Such subsets and memories are used to ``write'' and ``read'' the
information into and from the chain, and constitute areas
of {\em complete control} for the communicating parties
(namely Alice has total control on $A+C_A$, while Bob has
total control on $B+C_B$).

In the communication scenario we are considering here the spins of
the chain and Bob's memories $B$ are initially set up in the
$\hat{\sigma}_z$ ``down'' state $\ket{0} \equiv \ket{\downarrow}$
(analogously we indicate the $\hat{\sigma}_z$ ``up'' vector
with $\ket{1} \equiv \ket{\uparrow}$).
On the other hand Alice's memories $A$ are in some
(possibly entangled) input states $|\Psi \rangle_A$ 
which encode the information she wants to transmit.
At time $t=0$ the global state of the composite system $A+C+B$ is thus
\beq
\vert \Psi \rangle_A \otimes \vert {\bf 0} \rangle_C
\otimes \vert {\bf 0} \rangle_B \;, \label{statoin}
\eeq
(which is an eigenstate of the Hamiltonian of the chain~$\hat{\cal H}$).
Ideally, Alice's and Bob's goal is to transform~\eqref{statoin} into the state 
\beq
\vert {\bf 0} \rangle_A
\otimes \vert {\bf 0} \rangle_C \otimes \vert \Psi \rangle_B  \;, \label{statoout}
\eeq
by performing local operations on $A + C_A$ and $B+C_B$ and by
cleverly exploiting the transport properties of the chain free evolution.
In this context the efficiency of the communication can be 
characterized through the transmission rate $r$ of the protocol.
This is an asymptotic quantity which describes the maximum number
of qubits one can transfer per unit of time with average fidelity
converging to $1$ in the limit of large transmission
times (see for instance~\cite{MEMORY}).
In the present paper however, we will mostly focus on protocols for which 
the average time ${T}$ it takes to pass from~\eqref{statoin} to~\eqref{statoout}
is a finite quantity. In this case $r$ is given by
\begin{eqnarray}
r = M/{T} \;, \label{trans}
\end{eqnarray}
where $M$ is the number of qubits encoded into the states $|\Psi\rangle_A$
of Eq.~\eqref{statoin}.
In what follows ${T}$ will be computed by considering only the 
spin chain free evolution, thus neglecting the time intervals  
employed by Alice and Bob to perform their local operations. 
This is legitimate by the fact that in the model the only 
dynamical constraint is imposed by $\hat{\cal H}$.


Even in the ideal case when the coupling with the environment and the presence of
imperfections is neglected, the evaluation of the transmission rate of 
Eq.~\eqref{trans} is typically complicated by the dispersive free evolution of the
chain~\cite{OSBORNE,vgdaniel,bose03,daniel05,daniel05B,daniel05C,vittorio05}.
To better understand this point it is sufficient to focus on the case
in which $|\Psi\rangle_{A}$ is a separable vector of the form
\beq
|\Psi\rangle_A = | \cdots  \; \psi_3\; \psi_{2} \;\psi_{1} \rangle_A
\label{separabile}
\eeq
where the $j$-th memory element of Alice's memory is initialized in
$\vert \psi_j \rangle_{A_j} \equiv \alpha_j |0\rangle_{A_j} 
+ \beta_{j} |1\rangle_{A_j}$.
Suppose now that
Alice starts the transfer protocol at time $t=0$ by coupling
her first memory element $A_1$ with the chain spin $C_A$ 
through an instantaneous SWAP gate~\cite{NIELSEN} $\hat{S}_A (1)$.
This resets the memory element $A_1$ to $|0\rangle_{A_1}$ while ``copying''
its initial state into the first chain element, i.e.
\begin{eqnarray} \nonumber 
 \vert  {\bf \Psi \, 0 \, 0} \rangle_{ACB} &\rightarrow& 
|\cdots\;  \psi_{3} \; 
\psi_{2} \; 0\rangle_{A} \otimes \; \big( \;\alpha_1\; | 0 0 \cdots 00\rangle_C \\
&& + \beta_1 \;  |1 0\cdots 00\rangle_C \; \big) \otimes |{\bf 0}\rangle_B \; .
\label{statoin1}
\end{eqnarray}
The system then evolves freely for a time interval $\tau$
in order to allow the ``perturbation'' $\psi_1$ introduced locally by Alice
in $C_A$ to spread along the whole chain.
Since the Hamiltonian $\hat{\cal H}$ commutes with $\hat{\cal S}_z$,
the state~\eqref{statoin1} becomes~\cite{bose03}:
\begin{eqnarray}
|\cdots \psi_{3}
\psi_{2} 0\rangle_{A} \otimes
\left( \, \alpha_1 | {\bf 0} \rangle_C +
\beta_1 \sum_{n=1}^N \gamma_{1 n}(\tau) | {\bf n} \rangle_C \, \right) \otimes  
|{\bf 0}\rangle_B \;, \nonumber
\end{eqnarray}
where $|{\bf n}\rangle_C$ represents the state of the chain having
all spins down but the $n$-th, and where
\begin{eqnarray}
\gamma_{1 n}(\tau) =
{_C \langle {\bf n} |} e^{-i \hat{\cal H}\tau/{\hbar}} | {\bf 1} \rangle_C
\label{ampiezze}
\end{eqnarray}
is the probability amplitude of finding the spin up in the $n$-th chain location.
By applying the instantaneous SWAP gate $\hat{S}_B(1)$ which couples 
his first memory $B_1$ and the last chain element $C_B$,
Bob has now a chance to transfer Alice's information into $B$.
If the chain Hamiltonian $\hat{\Ham}$ is engineered such that there exists a certain
time $\tau_*$ at which the amplitude $\gamma_{1N} (\tau_*)$ is unitary~\cite{DATTA}
(i.e. $\gamma_{1N} (\tau_*) = e^{i \varphi_*}$), then the excitation
sent by Alice has been perfectly traveled to the spin $C_B$.
Bob can thus safely transfer the exact state $\psi_1$ into $B_1$
with a simple swap operation, followed by a proper phase shift gate on $B_1$
to compensate $e^{i \varphi_*}$.
This process can be iterated to the remaining memories $A$:
at the $j$-th run Alice will move the memory $A_j$ into the chain by means
of the SWAP $\hat{S}_{A}(j)$  which couples $A_j$ with $C_A$ while,
after a time interval ${\tau_*}$, Bob will extract it
from $C$  by applying the SWAP $\hat{S}_B(j)$ which couples  $B_j$ with $C_B$.
Assuming perfect timing, the scheme guarantees the transfer of one qubit
every ${\tau_*}$ seconds, yielding a rate~\eqref{trans} equal to 
$1/{\tau_*}$. 

Unfortunately for a generic Hamiltonian $\hat{\Ham}$ and time $\tau$
the amplitude $\gamma_{1N}(\tau)$ is not unitary; in this case Bob's SWAP will not
succeed in perfectly extracting Alice's information $\psi_1$ out of the chain.
The excitation which codifies $\psi_1$, that has been previously put into the chain,
is in general spread out over all the sites of the chain.
Therefore, at each run only a fraction of Alice's information is transferred in $B$:
the rest remains into the chain and has a chance of interfering with the
subsequent operations of the communicating parties.
In particular, every time Alice couples her memories with the chain,
there is a finite probability that part of the information which
was previously injected into $C$, will re-enter into $A$.
In this case she will never send it back through the chain, so that Bob
will never be able to reconstruct the state with perfect fidelity.
The net result is the arising of memory effects~\cite{MEMORY,MEMORY1}
in the communication which require a proper handling.

\section{Two-spins chain channel}\label{sec:due}

In this section we focus on the simplest non trivial spin channel model.
It is given by a chain $C$ of only $N=2$ spins, the first being
controlled by Alice and the second by Bob.
We will see that, despite its simplicity, the model retains sufficient
structure to permit the analysis of memory effects. 
In particular it will allow us to compare the transmission rates
of protocols which exploit memory effects with protocols which do not.

\subsection{Plain scheme}\label{sec:brute}

We begin by considering a communication scheme where,
every $\tau$ seconds, Alice and Bob simultaneously~\cite{NOTA}
perform sequences of SWAPs operations which couple the $A$ memories
with $C_A$ and the $B$ memories with $C_B$ (namely, at the $j$-th step
they both apply $\hat{S}_A(j)$ and $\hat{S}_B(j-1)$, respectively).
This is the simplest approach, in which the communicating parties 
try to squeeze their messages through the chain by repetitively
tempering with it, without taking into account its internal dynamics.
After $m$ steps the global state of the system is described by the vector
\begin{eqnarray}
\hat{\mathcal{W}}_m \vert \Psi {\bf \, 0 \, 0} \rangle_{ACB}
\label{statoglobale}\;,
\end{eqnarray}
where $\hat{\mathcal{W}}_m$ is the unitary transformation
\begin{eqnarray}
\hat{\mathcal{W}}_m &=&  \hat{S}_B (m-1) \, \hat{S}_A (m) \,
\hat{U} \cdots \label{doppiaw}\\
&& \cdots
\hat{S}_B (2) \, \hat{S}_A (3) \, \hat{U}  \:
\hat{S}_B (1) \, \hat{S}_A (2) \, \hat{U}  \: \hat{S}_A (1)\;,
\nonumber 
\end{eqnarray}
with $\hat{U} \equiv e^{ -i \hat{\cal H} \tau/\hbar}$.
Consequently the reduced density matrix of Bob's memories
can be expressed as
\begin{eqnarray} 
\rho_B(m) = \mbox{Tr}_{AC} \Big[ \;
\hat{\mathcal{W}}_m \; \Big( \; | \Psi {\bf \, 0 \, 0} \rangle_{ACB} 
\langle   \Psi {\bf \, 0 \, 0}| \; \Big) \; \hat{\mathcal{W}}_m^\dag \; \Big]
\label{eq:red}\;.
\end{eqnarray}
Despite the complexity of the correlations introduced
by the concatenated SWAPs, for $N=2$ Eq.~\eqref{eq:red} can
be reduced  to a tensor product form 
${\cal D}_\eta^{\otimes m}(| \Psi\rangle 
\langle \Psi |)$ (with ${\cal D}_\eta$ being the single 
qubit amplitude damping channel 
map~\cite{NIELSEN,bose03,vittorio05}) for which standard 
memoryless quantum channel analysis~\cite{SHOR} can  be used
to compute the rate~\eqref{trans};
this is not true in general for $N\neq 2$.
Suppose in fact that Alice's memories $A$ have been prepared in
the separable state of the form~\eqref{separabile}.
One can then verify that Eq.~\eqref{statoglobale} yields
\begin{figure}[t]
  \includegraphics[scale=0.8]{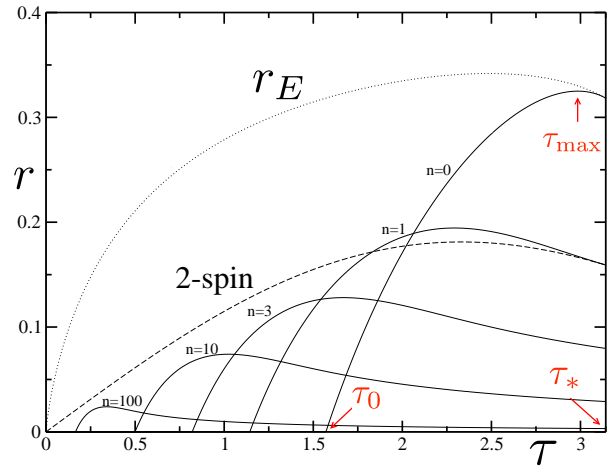}
  \caption{Transmission rates for the two-qubit channel as a function of $\tau$.
    The free evolution of the channel is driven by the Hamiltonian
    in Eq.~\eqref{eq:hamevol2qubit}, in which we set $J=1/4$.
    Continuous and dashed lines indicate the quantum transmission
    rate respectively for a standard, and a two-spin qubit encoding.
    The various continuous lines are for different numbers $n$ of Bob's
    intermediate swaps after each Alice's and Bob's simultaneous swap.
    For reference we also plot the rate $r_{E}$ associated to the 
    entanglement-assisted quantum transmission for the standard encoding
    (dotted line), which, analogously to Eq.~\eqref{VVV},
    can be computed by $r_{E} \equiv {Q_E(\eta)}/{\tau}$,
    with $Q_E (\eta)$ being the entanglement-assisted capacity
    of a single amplitude damping channel~\cite{vittorio05}, i.e.
    $Q_E (\eta) =(1/2) \max_{p \in [0,1]}{ 
      H_2 (p) + H_2 ( \eta p ) - H_2 \big( (1-\eta) p \big) }$.}
  \label{fig:ratesJ025}
\end{figure}
\barr
\nonumber {}
{} & \displaystyle \sum_{j_1 \ldots j_{m} = 0}^{1} &
\bar{\beta}_1^{j_m} \ldots \bar{\beta}_{n}^{j_{1}}
\vert j_m \ldots j_{1} \rangle_A \otimes
\vert \psi_{m+1} 0 \rangle_C \\
\label{eq:stato2spin}
 & & \otimes \,
\vert \tilde{\psi}_1^{(j_1)} \ldots \tilde{\psi}_{m}^{(j_m)} \rangle_B
\earr
where $\bar{\beta}_k = \beta_k \; \gamma_{11}(\tau) $, and
$$
\vert \tilde{\psi}_k^{(j_k)} \rangle = \left\{
\begin{array}{ll} \vert 0 \rangle \quad & \textrm{if} \quad j_k = 1 \\
\vert \psi^{'}_k \rangle \equiv \alpha_k \vert 0 \rangle +
\beta_k \gamma_{12}(\tau) \vert 1 \rangle \quad & 
\textrm{if} \quad j_k = 0 \end{array}\;, \right.
$$
with $\gamma_{11}(\tau)$ and $\gamma_{12}(\tau)$ defined as in Eq.~\eqref{ampiezze}
and satisfying the constraint
$\vert \gamma_{11} (\tau) \vert^2 + \vert \gamma_{12} (\tau) \vert^2 =1$.
We notice that, at each time after having applied the two SWAP operations,
the chain is always disentangled from Alice and Bob memories
(see Eq.~\eqref{eq:stato2spin}).
This permits to express Eq.~\eqref{eq:red} as a product of $m$
amplitude damping channels~\cite{NIELSEN,bose03,vittorio05} 
$\mathcal{D}_\eta$, with quantum efficiency
$\eta = \vert \gamma_{12}(\tau) \vert^{2}$ equal to the transfer probability
of one excitation from the first to the second spin of the chain.
Indeed, neglecting a phase shift component that can always be compensated
by Bob through a local operation on $B$, one has 
\begin{eqnarray}
\rho_B (m) &=& \bigotimes_{k=1}^{m} \left(
\vert \bar{\beta}_k \vert^{2} \, \vert 0 \rangle \langle 0 \vert +
\vert \psi_{k}^{'} \rangle \langle \psi_k^{'} \vert \right)
\label{eq:map2qubit} \\
&=& \bigotimes_{k=1}^m \mathcal{D}_{\eta}
\left(|\psi_k \rangle_{A_k}\langle \psi_k |
\right) = \mathcal{D}_{\eta}^{\otimes m} 
\left(|\Psi \rangle_{A}\langle \Psi |
\right)\;.
\nonumber \end{eqnarray}
Exploiting the linearity of Eq.~\eqref{eq:red} this identity can then be
generalized to all (non necessarily separable) input states $|\Psi\rangle_A$.

The transmission rate~\eqref{trans} associated with the protocol
in Eq.~\eqref{doppiaw} can be now easily computed by considering
the quantum channel capacity~\cite{SHOR} of the memoryless
channel map ${\cal D}_\eta$. This has been derived in
Ref.~\cite{vittorio05}: it is null for $\eta\leqslant 0.5$ and equal to
\beq
Q (\eta) = \max_{p \in [0,1]} \left\{ H_2 ( \eta p ) -
H_2 \left( (1-\eta) p \right) \right\} \label{NNtheq}
\eeq
otherwise (here $H_2 (x)= -x \log_2 x - (1-x) \log_2 (1-x)$ is the binary entropy
function). Equation~\eqref{NNtheq} gives the maximum number of qubits
which can be reliably transmitted  per use of the channel ${\cal D}_\eta$
in the asymptotic limit of $m \gg 1$ uses.
Considering that in a time interval ${T}= m \, \tau$ the protocol~\eqref{eq:red}
accounts for $m$ uses of the ${\cal D}_\eta$, we can estimate its rate as follows:
\beq
r \equiv \lim_{m \to \infty} \frac{m \; Q(\eta)}{m \, \tau}
= \frac{Q(\eta)}{\tau} \, . \label{VVV}
\eeq
It should be stressed that the possibility of achieving the rate~\eqref{VVV}
relays on the identification of an optimal encoding space~\cite{SHOR}
which, in the general case, requires infinitely many uses
of the map ${\cal D}_\eta$ (i.e. infinitely long transmission time).
In this respect Eq.~\eqref{VVV} should be considered more 
as an indication of the efficiency of the protocol~\eqref{eq:red}
rather than a realistic communication rate of the chain.

In order to provide an explicit expression for the rates~\eqref{VVV},
we consider the $\hat{S}_z$-preserving spin chain Hamiltonian of the form
\beq
\hat{\Ham} = \hbar \; J (\hat{\sigma}^x_1 \hat{\sigma}^x_2 + \hat{\sigma}^y_1
\hat{\sigma}^y_2 ) + \hbar \; \Delta \hat{\sigma}^z_1 \hat{\sigma}^z_2 \;,
\label{eq:hamevol2qubit} \eeq
for which the excitation transfer amplitude
is just a sinusoidal periodic function of $\tau$ of period
 $\pi / (2J)$, i.e.
\beq
\eta = |\gamma_{12}(\tau)|^2 = 
\sin^2 (2 J \, \tau ) \;. \label{sin}
\eeq
For $\tau = {\tau_*} \equiv \pi/(4 J)$ the free evolution operates a SWAP
between the spins, thus achieving perfect transmission of a generic quantum state.
Correspondingly in this case the quantum capacity~\eqref{NNtheq} of the channel
is optimal and equal to one while the rate~\eqref{VVV} is $1/{\tau_*}$.
Given the periodicity of Eq.~\eqref{sin}, for $\tau> {\tau_*}$
the rate~\eqref{VVV} can never be higher than this quantity.
However, there exists a value $\tau_{\mathrm{max}} < {\tau_*}$
such that $r (\tau_{\mathrm{max}}) > r ({\tau_*})$ (see
Fig.~\ref{fig:ratesJ025}): i.e. ${\tau_*}$ is not the optimal 
time transfer for the plain scheme~\eqref{eq:red}. 
Notice also that the quantum capacity of the amplitude damping channel
is strictly zero for $\eta \leqslant \eta_c = 0.5$, thus meaning that
for $\tau \leqslant \tau_0 \equiv \pi/8 J$, the channel does not transmit
any quantum information.

In order to reduce the time $\tau_0$, a slightly different version
of the protocol can be implemented, in which we suppose that Bob has
at his disposal $n$ additional memories for each qubit that Alice sends.
The protocol goes on exactly as before, except that, after each double swap,
(performed from both Alice and Bob) Bob runs $n$ additional SWAP
operations at regular time intervals $\tau$.
The unitary transformation in Eq.~\eqref{doppiaw} then modifies into:
\begin{eqnarray} 
\nonumber \hat{\mathcal{W}}_m &=&
\big[ \hat{S}_B (m_n) \hat{U} \cdots \hat{S}_B (m_1) \hat{U} \big]
\,  \hat{S}_A (m) \, \hat{U} \cdots \\ \nonumber && \cdots
\big[ \hat{S}_B (2_n) \hat{U} \cdots \hat{S}_B (2_1) \hat{U} \big]
\,  \hat{S}_A (2) \, \hat{U} \cdot \\ &&
\cdot \big[ \hat{S}_B (1_n) \hat{U} \cdots \hat{S}_B (1_1) \hat{U} \big]
\,  \hat{S}_A (1)  \;.
\label{doppiaw2}
\end{eqnarray}
In this way Bob can enhance the transfer fidelity~\cite{vgdaniel},
at the price that both time and memory requirements are increased
by a factor $n$ for each qubit sent by Alice.
The capacity of such a channel can be evaluated exactly as before,
except that the quantum efficiency $\eta$ is now dependent of $n$,
and it is given by:
\beq
\eta_n \equiv 
\sum_{k=0}^n \vert \gamma_{11} (\tau) \vert^{2k} \vert \gamma_{12} (\tau) \vert^2 
= 1 - (1 - \eta)^{n+1} \;,
\eeq
where $\eta=\vert \gamma_{12} (\tau) \vert^2$ is the quantum efficiency for $n=0$.
In Fig.~\ref{fig:ratesJ025} we plotted the corresponding 
quantum transmission rates as a function of the time
between two successive swaps $\tau$ for different values of $n$.
Notice that the time $\tau_0$ is reduced, as one increases $n$.

\subsection{Exploiting the internal dynamics of the chain} 
\label{sec:mult}

In this section we show how memory effects induced by the free evolution
of the chain can be exploited in order to simplify encoding and decoding
procedures. In particular, differently from the cases discussed in the
previous section, the schemes analyzed here allow one to achieve optimal 
transmission rate by encoding the information in only a finite number
of memory elements. 

The simplest version of these new classes of protocols is a variation
of the ``dual-rail encoding'' of Ref.~\cite{daniel05}.
The idea is to assume that
Alice uses her first two memory qubits (i.e. $A_1$ and $A_2$) to 
codify a single information qubit $|\psi\rangle = \alpha |0\rangle +
\beta |1\rangle$, while keeping the third memory element into the reference
state $|0\rangle_{A_3}$, i.e. following the notation of Eq.~\eqref{separabile}
\begin{eqnarray}
|\psi\rangle &\rightarrow& 
\alpha |010\rangle_A + \beta |001\rangle_A
\;. \nonumber
\end{eqnarray}
As in the plain scheme, every $\tau$ seconds Alice and Bob are then
required to perform a sequence of SWAPs gates between their memories
and the chain. In this case however, we will show that
after the second SWAP by Bob (i.e. after the third SWAP by Alice) 
a simple magnetization measurement on their memories allow
both the communicating parties to establish, independently,
whether the state $\vert \psi \rangle$ has been exactly transmitted to Bob,
or it has returned to Alice's memory.
Indeed, assume that at $t=0$ the global system is in the state
\beq
\big( \alpha \vert 010 \rangle_A + \beta \vert 001 \rangle_A \big)
\otimes \vert 0 0 \rangle_C \otimes \vert 00 \rangle_B \;.
\eeq
After the first two SWAPs of Bob and the first three SWAPs of Alice
(i.e. after $2 \tau$ seconds from the beginning of the transmission),
it is transformed into a superposition where with probability
$|\gamma_{11}(\tau)|^2$ the information has been returned into $A$
(encoded in $A_3 A_2$), while with probability $|\gamma_{12}(\tau)|^2$
the information has been moved into $B$ (encoded in $B_1B_2$), i.e.
\begin{eqnarray}
&&\gamma_{11}(\tau) \; 
\big( \, \alpha \vert 100\rangle_A + \beta | 010\rangle_A \, \big) \otimes
\vert 00 \rangle_C \otimes \vert 00 \rangle_B \label{eq:double_spin} \\
&&+\; \gamma_{12}(\tau) \; \vert 000 \rangle_A
\otimes \vert 00 \rangle_C \otimes \big( \,\alpha \vert 10 \rangle_B
+ \beta | 01 \rangle_B \, \big) \;.
\nonumber 
 \end{eqnarray}
These two possibilities can be distinguished by Alice and Bob by performing 
independent magnetization measurements on their respective memories $B$ and $A$.
For instance, the first possibility (i.e. information in $A$)
will yield, respectively, the outcome $0$ (null total magnetization of $B$)
and $1/2$ (a single spin up in $A$) for Bob and Alice measurements.
Analogously, when the information is in $B$ the measurements
will yield, respectively, the outcome $1/2$ and $0$.
In the latter case the communicating parties can proceed by
sending another qubit (encoded by Alice in $A_6 A_5 A_4$ and
received by Bob in $B_3 B_4$), while in the former case, first the
information is (locally) moved back from $A_3A_2$ into $A_2A_1$ and the
protocol is repeated until Bob is certain to receive the state.
The iteration of this procedure is trivial.

\begin{figure}[t]
  \includegraphics[scale=0.7]{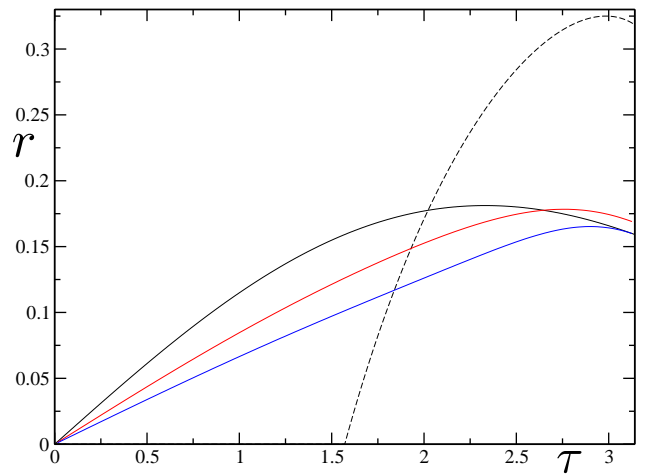}
  \caption{Quantum transmission rates for a two-qubit channel driven by the
    Hamiltonian in Eq.~\eqref{eq:hamevol2qubit} with $J=1/4$ for different
    qubit encodings: standard encoding (one spin per qubit, dashed black line),
    one qubit in two spins (black line), two excitations in three spins
    (red line), three excitations in four spins (blue line).}
  \label{fig:Mult_spin}
\end{figure}

To compute the transmission rate of this communication scheme
we note that the probability that Bob will receive Alice's
state $|\psi\rangle$ exactly at the $j$-th iteration of the protocol is
$P_j = \vert \gamma_{12}(\tau) \vert^2 \; 
(1 - \vert \gamma_{12}(\tau) \vert^2 )^{j-1}$.
The average time required to transfer the qubit is then:
\beq
T = \sum_{j=1}^{+ \infty} 2 \tau j P_j 
= \frac{2 \tau}{|\gamma_{12}(\tau)|^2} \;, \eeq
from which we get
\beq
r = \frac{1}{T} = \frac{\vert \gamma_{12}(\tau) \vert^2}{2 \tau}
\label{eq:1ecc} \;.\eeq
Assuming that the transferring spin chain $C$ is described by the
Hamiltonian in Eq.~\eqref{eq:hamevol2qubit}, this expression has been plotted
in Fig.~\ref{fig:ratesJ025} (dashed line) for a comparison with the protocols
of the previous section.
Notice that, for $\tau \leqslant \tau_0$, contrary to the standard
plain encoding, the transmission rate is not zero; the maximal transfer rate
however is achieved with a standard encoding.

\begin{figure}[t]
  \includegraphics[scale=0.6]{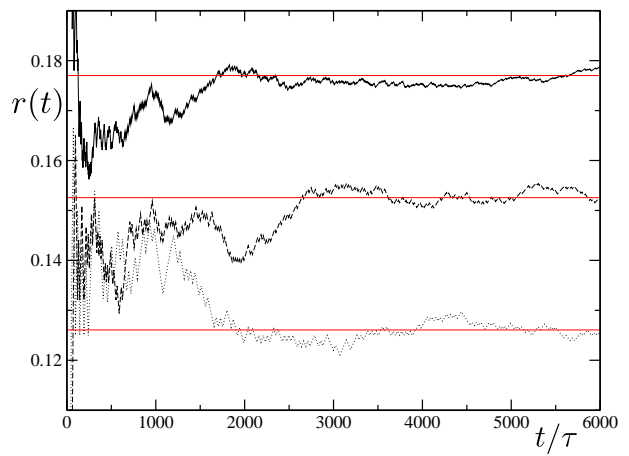}
  \caption{Instantaneous quantum transmission rates for a two-qubit channel
    described by the Hamiltonian in Eq.~\eqref{eq:hamevol2qubit} with $J=1/4$
    for $\tau = 2$. The various curves correspond to different qubit encodings:
    one qubit in two spins (continuous line), two excitations in three spins
    (dashed line), three excitations in four spins (dotted line).
    Straight red lines correspond to the rates obtained after averaging over an
    infinite time, which are evaluated analytically
    from Eqs.~\eqref{eq:1eccM},~\eqref{eq:2ecc} and similar.}
  \label{fig:MC_tau2}
\end{figure}

Alice and Bob can use slightly more complicated types of encodings,
in order to optimize the transfer rate.
For instance Alice can fix the number $E$ of excitations she employs
to codify her input qubit messages in $\mathcal{N}$ spins of $A$.
The case discussed before corresponds to $E=1, \: \mathcal{N}=2$;
the generalization to a generic number of spins, with $E=1$ fixed, is trivial:
Alice can send a number $\log_2 (\mathcal{N})$ of qubits, provided she employs
$\mathcal{N}+1$ memories (the extra memory play the same role of $A_3$ in the simple
version of the scheme). The protocol then proceeds exactly as before,
where Bob swaps on his $\mathcal{N}$-states memory.
He then has to measure the magnetization at every time interval $\mathcal{N} \tau$.
The success probabilities $P_j$ are the same as before,
while the transfer rate is then given by:
\beq
r = \frac{\vert \gamma_{12}(\tau) \vert^2 \;\log_2 \mathcal{N}}{\mathcal{N} \tau}
\label{eq:1eccM}\;.
\eeq
The case $E=2$ is slightly more complicated by the fact that,
after a time $\mathcal{N} \tau$, Bob can measure two excitations
with probability $\vert \gamma_{12} (\tau) \vert^4$
(in that case he has perfectly received the state), no excitations with
probability $\vert \gamma_{11} (\tau) \vert^4$ (the state has perfectly returned
to Alice, therefore they have to restart the protocol), or one excitation.
In this last case, only one excitation is returned to Alice and she has then 
to retransmit it, by using the same procedure for $E=1$ described before.
It can be shown that the transfer rate for the case $E=2$ is given by:
\beq
r = \frac{\log_2 {\mathcal{N} \choose 2}}{\mathcal{N} \tau} \cdot
\frac{ (1- \vert \gamma_{11} (\tau) \vert^4)^2}
{1- \vert \gamma_{11} (\tau) \vert^4 + 2 \vert \gamma_{11} (\tau) \vert^2
- 2 \vert \gamma_{11} (\tau) \vert^6} \, .
\label{eq:2ecc} \eeq
Similar expressions for the transfer rate with higher $E$ can be obtained.
The only difference is that an increasing number
of possibilities appears: after a time $\mathcal{N} \tau$, Bob can receive a number
of excitations $E_B \leqslant E$, 
consequently a number $E_A = E -E_B$ of excitations return to Alice.
According to the value of $E_A$, she then has to apply a sub-protocol for
the transfer of $E_A$ excitations, with $E_A \leqslant E$. 
This procedure has to be iterated until $E_B = E$.

In Fig.~\ref{fig:Mult_spin} we show the theoretical values of the transmission rates
in a two-qubit channel for different values of $E$ and $\mathcal{N}$, as a function
of the time $\tau$ (continuous lines); the asymptotic rate for the
standard encoding, Eq.~\eqref{VVV}, is also shown for reference.
Moreover we have explicitly simulated these types of communication protocols
between Alice and Bob with a standard Monte Carlo numerical technique.
To this end, an instantaneous transmission rate $r ({t})$ can be defined
as the ratio between the number of transmitted qubits $M$ until time ${t}$ 
and the actual transmission time ${t}$.
The value of $M$ has been evaluated stochastically, following the theoretical
probability distributions $P_j$ of the state transfer.
In Fig.~\ref{fig:MC_tau2} we explicitly show the dependence of such computed
instantaneous transmission rate as a function of the elapsed time $t$,
for different types of encodings; notice that, by definition
of transmission rate~\eqref{trans}, the instantaneous transfer rate
$r ({t})$ correctly converges to the asymptotic value
given by Eqs.~\eqref{eq:1eccM},~\eqref{eq:2ecc} and similar (straight red lines).

\section{Dual-rail channel}\label{sec:dualrail}

In the previous section we analyzed the
simplest spin chain model ($N=2$). The results we obtained
were indicative of the possibility of exploiting memory effects
to devise better communication procedures (e.g. having simpler encoding
and decoding protocols). These results also showed that transmission rates
could easily be computed also in the presence of such effects.
In this section we would like to derive a lower bound for the maximum
achievable transmission rate that can be reached in the case
of an arbitrarily long spin chain. 
This is not a simple task~\cite{MEMORY,MEMORY1}, due to the presence
of the memory correlations in the evolution of the spins chain.
The point is that, at present, given a single spin chain of $N>2$ elements,
we do not have communication schemes which permit Alice and Bob
to verify independently that the transferring of a signal succeeded, allowing,
on one hand, to move to the transmission of the next one, while,
on the other hand, determining the average qubit transmission time $T$.
A simple way to address  these issues, is to  consider the case in which 
the channel $C$ is composed by three identical uncoupled spin $1/2$ chains,
each of them governed by an Hamiltonian $\hat{\Ham}$ that conserves
the total magnetization.

\begin{figure}[t]
  \includegraphics[scale=0.67]{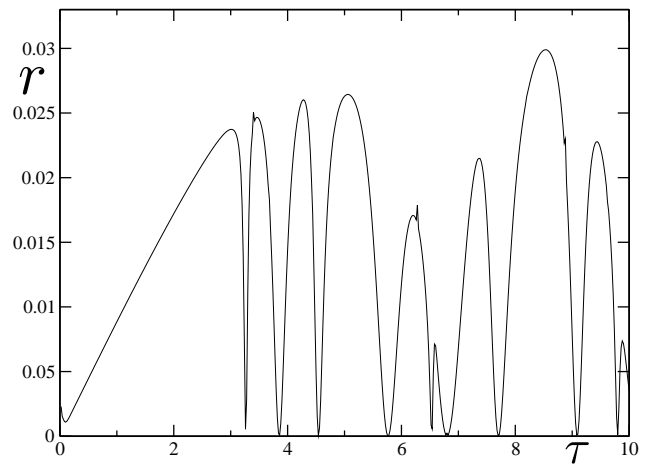}
  \caption{Quantum transmission rates as a function of $\tau$ for a	
    dual-rail channel with classical feed-back side line for a chain of
    $N=8$ spins. Free evolution of the chains is governed by
    an Heisenberg chain with $J=1/4$.
    Data have been obtained numerically by truncating the series
    in Eq.~(\ref{serie}) at $k_{max} = 10^5$.}
    \label{fig:2Rail_N8}
\end{figure}

As before, we suppose that Alice has access to the leftmost spin of each of 
the three chains, while Bob can manipulate the spin at the opposite end of the chains.
We also assume that, at time $t=0$, all the chains are set up in the
ferromagnetic ground state
$\ket{{\bf 0}}^{(i)} \equiv \vert 0_1 \ldots 0_N \rangle^{(i)}$
(where $i=\{ 1,2,3 \}$ is the index that labels the chain).
The communication strategy we want to analyze is the following.
Alice use the chains $1$ and $2$ to transfer her first message to Bob
by means of a dual-rail encoding~\cite{daniel05}. Since this is a ``conclusive''
strategy, it allows Bob to know exactly at what instant Alice's message
has been loaded in his memory. When this happens, he will use the third chain
to signal back to Alice that he is ready to receive a new qubit of information
(e.g. he does so by sending a spin up message to Alice).
The whole procedure is then reiterated for the transmission of the
second Alice's message.

To see how this works in details let us first consider a simplified version
of the above scheme, where the feed-back message by Bob is transmitted
to Alice through a side classical communication line (e.g. a telephone line).
In this case we need only to consider the information transfer
along the spin chains $1$ and $2$ from Alice to Bob.
Assume that the first message Alice wants to transmit is the qubit
$\ket{\psi}=\alpha|0\rangle + \beta \ket{1}$.
The chains $1$ and $2$ are then prepared into the following superposition:
\beq
\ket{ {\bf s} (1) } \equiv \alpha \ket{{\bf 0}}^{(1)}
\otimes \ket{{\bf 1}}^{(2)} +
\beta \ket{{\bf 1}}^{(1)} \otimes \ket{{\bf 0}}^{(2)} \, .
\label{eq:dual_init}
\eeq
The system is then let freely evolve, such that the excitation in
Eq.~\eqref{eq:dual_init} will propagate along the two chains:
\beq
\ket{ {\bf s} (1) } \stackrel{\hat{U}(\tau)}{\longrightarrow}
\sum_{n=1}^{N} \gamma_{1n} (\tau) \ket{ {\bf s} (n) } , \label{eccoqui}
\eeq
where $\gamma_{ij} (\tau)$ is the same as in Eq.~\eqref{ampiezze}.
Following Ref.~\cite{daniel05}, at regular time intervals $\tau$ Bob 
performs a magnetization measurement
on the last spins of the chain $1$ and $2$, 
in order to check if the state $\ket{\psi}$ has traveled
to him. In the meantime Alice does nothing and waits until
she receives Bob's ``OK'' feed-back message on the phone.
At the first Bob's measurement, which happens after a time $\tau$, if he measures
a non-zero magnetization, he concludes that the qubit $\ket{\psi}$ is located
on the last spins of the chain: therefore he can safely SWAP it into his memory $B$. 
According to Eq.~\eqref{eccoqui}, such event happens with probability 
$\pi_1 \equiv \vert \gamma_{1N} (\tau) \vert^2$.
In this case he communicates to Alice via the classical channel the success
of information transfer, and she will proceed by sending another qubit
through the chains $1$ and $2$ following the same procedure.
\begin{figure}[t]
    \includegraphics[scale=0.55]{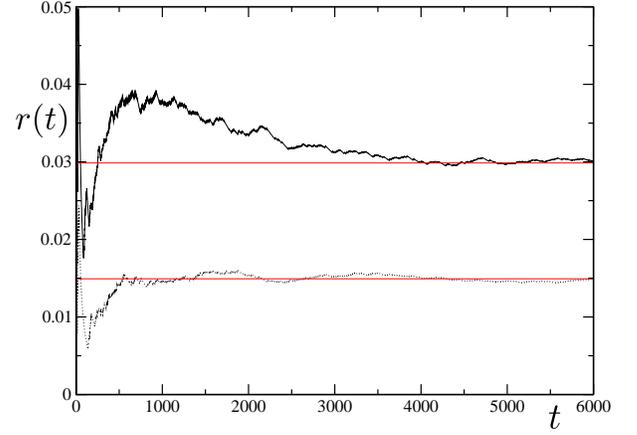}
    \caption{Instantaneous quantum transmission rates for a dual-rail channel
      with $N=8$ spins, for $\tau =8.5$; $J=1/4$.
      Black curves shows the results of a Monte Carlo simulation of the protocol,
      straight red lines indicate the theoretical values of the rate,
      obtained after averaging over an infinite time.
      Upper curves correspond to the case in which the transmission protocol is
      assisted from a backward classical communication channel; in the lower curves
      backward communication from Bob to Alice occurs via a single quantum spin chain,
      equal to the two forward communicating ones.}
    \label{fig:dyn+bck_N8_tau85}
\end{figure}
Vice-versa, if the first outcome of Bob's  measurement is zero, 
then he knows that the system has been projected in the state
\beq
\frac{1}{\sqrt{1-\pi_1}} \sum_{n=1}^{N-1} \gamma_{1n} (\tau)
\vert {\bf s} (n) \rangle \;,
\eeq
where Alice's qubit of information $|\psi\rangle$ is still contained
in the chains $1$ and $2$.
Bob has then another possibility to receive the state $\ket{\psi}$:
he can wait for another time $\tau$, before performing the second
magnetization measurement.
Just before the measurement, the system will be in the state:
\beq 
\frac{1}{\sqrt{1-\pi_1}} \sum_{n=1}^{N} \Big( \gamma_{1n} (2 \tau)
- \gamma_{Nn} (\tau) \gamma_{1N} (\tau) \Big) \vert {\bf s} (n) \rangle
\;.\eeq
Correspondingly 
Bob's probability to receive the qubit at the second measurement is then:
\beq
\pi_2 = \frac{1}{\pi_1} \vert \gamma_{1N} (2 \tau)
- \gamma_{NN} (\tau) \gamma_{1N} (\tau) \vert^2 \;.
\eeq
If the transfer has been still unsuccessful, then he can repeat this strategy,
until he is sure the state has been transferred.
After each time $k \tau$ he has a probability $\pi_k$ to receive the state
that can be obtained by simply iterating this scheme:
\beq
\pi_k = \left[ \prod_{j=1}^{k-1} \frac{1}{1-\pi_j} \right] \cdot \vert c_k \vert^2
\;,\eeq
where
\beq
c_k = \gamma_{1N} (k \tau) - \sum_{j=1}^{k-1} \gamma_{NN} (j \tau) \: c_{k-j} \;.
\eeq
The probability of having $k-1$ failures and a success at the $k$-th measurement
is thus expressed by:
\barr
\nonumber P(k) & = & \pi_k \cdot (1 - \pi_{k-1}) \cdot (1 - \pi_{k-2}) \cdot
\ldots \cdot (1 - \pi_{1}) \\ {} & = & \bigg{\vert} \gamma_{1N} (k \tau) -
\sum_{j=1}^{k-1} \gamma_{NN} (j \tau) \: c_{k-j} \bigg{\vert}^2 \;.
\label{NNNprob}
\earr
The total probability of success after $k$ steps is given by the sum
of all $P(j)$, with $j=1, \ldots, k$.
It can be shown that, under a very general hypothesis on the system
Hamiltonian $\hat{\Ham}$, the probability of success converges to 1
in the limit $k \to \infty$~\cite{daniel05B}.

By knowing all the probabilities~\eqref{NNNprob} it is possible to evaluate the
average time $T$ needed for the transfer of the first qubit from Alice
to Bob. Indeed, since $P(k)$ is exactly the transfer probability after $k$ steps,
and since each step takes $\tau$ seconds, we get
\beq
T = \sum_{k=1}^{+ \infty} k \, \tau \, P (k) \;.
\label{serie}\eeq
If we suppose that Bob can instantaneously communicate to Alice the
fact that he effectively received the qubit (for example via a classical
communication channel), and that, immediately after having known the
transfer success, she sends another qubit, we than obtain  
the transfer rate $r = 1/T$.
In Fig.~\ref{fig:2Rail_N8} we plot this quantity for a dual rail channel
composed of two identical isotropic spin $1/2$ Heisenberg chains of the form
$\hat{\Ham} = - \hbar \;J \;\sum_{j=1}^{N} \,
\hat{{\vec{\sigma}}}_j^{(i)} \, \hat{{\vec{\sigma}}}_{j+1}^{(i)}$ for
which 
the amplitudes $\gamma_{ij}(\tau)$ have been explicitly computed in 
Ref.~\cite{bose03}.
As discussed  at the beginning of the section, the requirement 
of a classical communication channel needed as a feedback from
Bob to Alice can be relaxed, provided that there is a third spin chain
connecting them. In this case, when Bob has received the qubit,
he puts an excitation in this chain. Alice then uses the same protocol for the
forward communication in order to receive it.
If the third chain is identical to the others, the average time required
for the forward and the backward communication are equal, therefore
the rate for the quantum communication halves.

In Fig.~\ref{fig:dyn+bck_N8_tau85} we show the results of a Monte Carlo simulation
of these dual-rail protocols: 
the instantaneous transmission rates as a function of time
in units of $\tau$ are plotted. We simulated both the case assisted with a
backward classical communication channel (upper curves), and the case in which
backward communication from Bob to Alice occurs via a third quantum spin chain,
equal to the two forward communicating ones (lower curves).

\begin{figure}[t]
    \includegraphics[scale=0.55]{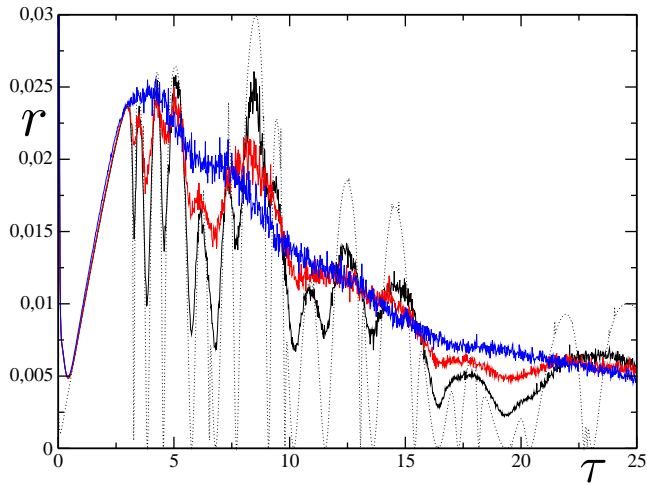}
    \caption{Quantum transmission rates for a dual-rail channel with $N=8$ spins,
      as a function of $\tau$. The various continuous curves are for different
      values of the random tilting time: $\epsilon=0.01$ (black), 0.02 (red),
      0.04 (blue curve). The dashed black line corresponds to the case
      of no tilting time ($\epsilon =0$).
      A truncation value of $k_{\mathrm{max}} =5000$ has been adopted here.}
    \label{fig:2Rail_N8_tvar}
\end{figure}

Finally we notice that, if the elapsed time between two successive Bob's measurements
$\tau$ is kept fixed, the transmission rate $r$ as a function of $\tau$ displays
a highly non monotonic behavior, which is typically unpredictable.
In particular there are some values of $\tau$ for which the
transmission rate suddenly drops to a value close to zero.
This is due to the sinusoidal quasi-periodic behavior of the amplitudes $\gamma$.
A possible strategy in order to reduce these singularities, would be that of
randomly varying the time interval between measurements, i.e. $\tau + \delta \tau$
(where $\delta \tau$ is randomly chosen in $(-\epsilon,\epsilon)$, and
$\epsilon$ indicates the strength of the tilting time).
Numerical results are shown in Fig.~\ref{fig:2Rail_N8_tvar}, where the various
curves refer to different values of $\epsilon$.
For small times $\tau$ the rate is independent of $\epsilon$, while at long times
its behavior approaches that of a power law $r \sim \tau^{-1}$.

\section{Conclusions} \label{sec:conclusions}

We have analyzed spin-chain communication protocols in which a single quantum
channel is used in order to admit multiple-qubit transfer in time between
two distant parties. 
Without using Hamiltonians engineered {\it ad hoc}, or complex encoding
and decoding operations, Bob is generally not able to perfectly recover
Alice's information, due to the dispersive free evolution of the chain.
Moreover, memory effects naturally arise in these types of protocols:
part of the information previously injected by Alice into the chain
typically interferes with subsequent data sent.
Nevertheless in some cases we are able to evaluate the transmission rates
even in presence of such memory effects.
In the case of a two-spin channel, despite the fact that the chain
can act as a simple swapper, thus making it possible to obtain a perfect
state transfer from Alice to Bob, we showed that the maximum achievable
transfer rate is not obtained in correspondence of perfect transfer.
Transmission rates along arbitrary long chains can be analyzed numerically
in the framework of the dual-rail protocol, thus permitting to establish
a lower bound for the maximum achievable rate.

\acknowledgments
We acknowledge useful discussions with D. Burgarth. 
This work was supported by EC through grant EUROSQIP and by MIUR- through PRIN.
The present work has been performed within the "Quantum Information" research
program of Centro di Ricerca Matematica ``Ennio De Giorgi'' of Scuola Normale Superiore.


\begin{thebibliography}{99}
\bibitem{NOTA}
  Simultaneity between Alice and Bob operations and the assumption
  of having uniform time intervals are not fundamental in our analysis:
  these hypotheses are considered only to simplify the problem
  and to compact the notation.
\end{thebibliography}
\end{document}